\documentstyle[12pt]{article}
\def\uv{($u$,$v$)} 
\def\as{$^{\prime\prime}$} 
\begin{document}
\twocolumn
\section*{Space Instrumentation: Imaging Interferometry}
Due to the degrading effects of the Earth's turbulent atmosphere, 
the spatial resolution achieved by ground-based optical astronomy is
limited to the extent of the {\sc seeing} disk - the image of a point source
(e.g. a single star), taken through the atmosphere.
The size of the Seeing disk is independent of telescope diameter, but changes
only with wavelength and climatic conditions - about 0.5\as\ at the best 
observing sites in good weather.
In the absence of atmospheric Seeing, e.g. in space, or with a perfect
{\sc adaptive optics} system, a telescope has a resolution 
defined by the {\sc diffraction limit}, which scales inversely with
telescope diameter: the larger the mirror, the sharper the image.
For example, the 
{\sc Hubble Space Telescope} with its 2.5~m aperture
has a diffraction limit of 0.05\as\ at a wavelength of 500~nm, ten 
times better than the Seeing limit, but still not good enough. Projects such
as resolving the immediate surroundings of black holes in external galaxies,
or imaging extra-solar planets, require telescope diameters of at least 100~m.
To manufacture and launch into space such large apertures is clearly out of 
reach of present day technology. Yet, images with such seemingly impossible
resolution are feasible. They can be obtained with a method called 
{\sc synthesis imaging} which can achieve basically unlimited
resolution without unreasonably large telescopes. In this article, I
discuss the principle of synthesis imaging and the advantages of
imaging interferometry from space. I also briefly describe
various projects planned for the near future 
which promise to revolutionize astronomical imaging.
While the general discussion of synthesis imaging is valid for the entire
electro-magnetic spectrum, the degree of difficulty of its realization 
varies dramatically with wavelength. Here, I will focus solely
on the optical regime, and ignore the rich and successful history 
of radio interferometry, which is discussed elsewhere in this volume. 
\subsection*{The Idea of Synthesis Imaging}
Figure 1a shows a simple star cluster as it would be seen by
a (hypothetical) diffraction-free 10~m telescope in space.
Imagine its primary mirror covered by an opaque mask with a set of 
holes along a line.
Obviously, most of the image information would be lost - but not all
of it! The image produced through such a mask shows a set of
{\sc interference} fringes - alternating stripes of high and low intensity 
(Fig. 1b). These fringes contain information about the structure 
of the object. Different rotation angles of the mask produce 
additional fringe patterns. All patterns can be recorded
and combined later to obtain the complete source image, as demonstrated in 
Figs. 1c and 1d. In fact, there is no need for the full 10~m telescope mirror:
the fringe patterns can be obtained with just a set (at least two) of small 
mirrors by combining their beams in a common focus. 
This is the elegant principle behind synthesis imaging: to 
mimic a large telescope with a number of smaller ones.

There is, however, one caveat which in the past has prevented the
successful realization of synthesis imaging at optical wavelengths.
In order to produce the all-important interference fringes,
the beams from the individual apertures must be combined {\it coherently}.
This means that the difference in their respective pathlengths
to the common focus must not exceed the {\sc coherence length} of the light. 
Combining the two signals coherently requires accurate 
control of the pathlengths between the two telescopes
and the beam combination optics. The enormous technological difficulties
associated with controlling optical pathlengths to the required levels 
of a few wavelengths are described elsewhere in this volume. 
\subsubsection*{How does it work?}
The mathematical framework of synthesis imaging is based on the theory
of wave diffraction. The basic result is summarized in
the {\sc van Cittert - Zernike theorem}:\\
{\it the complex degree of coherence is equal to the
normalized Fourier transform of the source intensity distribution\footnote{
Strictly speaking, this formulation of the van Cittert - Zernike theorem
is correct only for the case of an incoherent light source. Also,
both the source size and the baseline length must be much smaller 
than the distance to the source (the far-field approximation). 
However, astronomical sources at optical wavelengths always
meet all of these conditions.}.} \\
The degree of coherence is a complex quantity, the 
observables measured in practice are
its amplitude and phase. They can be derived from the interference pattern 
(the fringe) which appears on the detector when 
scanning the delay between the two beams.

An imaging interferometer measures the
degree of coherence across the \uv -plane. 
Because, in essence, this measures the Fourier transform
of the source structure, the vector between the two telescopes
- the baseline $\vec{B}$ -  is usually measured in the spatial frequency 
domain, i.e. in units of wavelengths, $u = B_x/\lambda$ and  $v = B_y/\lambda$.
The plane in which the source wavefront is sampled is therefore
called the \uv -plane. Different baseline vectors are
achieved by moving the individual telescopes with respect to the
source, thus sampling different \uv -points. According
to the van Cittert - Zernike theorem, this allows the full reconstruction
of the source structure via an inverse Fourier transform, 
{\it provided that the \uv -coverage is complete}. 
However, the \uv -coverage of any real interferometer will have gaps, 
resulting in ambiguity in 
the source reconstruction. The ring-like residuals in Fig. 1d are
one example for such ambiguities, also known as {\sc grating rings}.
The completeness of the \uv -coverage,
is therefore a critical design parameter for imaging interferometers. 
\subsubsection*{Spatial resolution}
To first order, the resolution of an imaging interferometer is defined
by the longest baseline. However, in contrast to the diffraction
limit of a filled aperture, the spatial resolution of a 
synthesis imaging observation is not uniquely defined.  
By varying the weighting of the data obtained at
different \uv -coordinates, or even
removing some baselines completely, one can put more or less emphasis on 
certain spatial frequencies in the reconstruction, and thus change the 
effective resolution, even after the data are taken.
This fact is regularly exploited at ground-based telescopes
in {\sc aperture masking} observations 
of bright sources for which collecting area is not an issue, 
but the highest possible resolution is required. 
\subsubsection*{Field of View}
The field of view (FOV) of an imaging interferometer
is limited by the coherence length of the light, 
$l=\lambda^2/\Delta\lambda$, where $\lambda$ and $\Delta\lambda$ are
the wavelength and the bandwidth of the observation.
While the path lengths from the two interferometer elements ideally
are equal in the center of the field, an angular separation
$\Theta$ from the center necessarily produces
a path difference $\Theta\cdot D$.
As long as this difference is small compared
to $l$, the fringe contrast will not be affected. 
Thus, $\Theta = < \frac{\lambda}{D}\cdot \frac{\lambda}{\Delta \lambda}$, or
$\Theta < R_{\rm spatial\it}\cdot R_{\rm spectral\it}$. 
The maximum FOV of an interferometer thus can be estimated from its 
spatial and spectral resolution.
In practice, however, the FOV is likely to be limited by the effects of 
the incomplete \uv -coverage (see Fig. 1). 
\subsection*{Synthesis Imaging in Space}
All that has been said so far about synthesis imaging applies equally 
to space- and ground-based interferometers. There are, however, important
differences between the two environments in a number of aspects: 

{\bf \uv -coverage:}
Most interferometer arrays will consist of relatively 
few elements because of cost constraints. In order to overcome the
intrinsically sparse \uv -coverage of such systems, many employ movable
apertures. In addition, all ground-based arrays use the Earth's rotation 
during the course of the night to sample many different \uv -points. 
The achievable \uv -coverage therefore depends both on the geographical
location of the array, and the source position with respect to 
the Earth's rotation axis.

In contrast, an interferometer in space cannot rely on the Earth's rotation. 
It therefore
must have movable apertures to fill the \uv -plane. On the other hand,
the \uv -coverage can be chosen freely. This allows a uniform resolution 
which can be valuable in survey programs which study the statistical
properties of large numbers of similar objects.

{\bf Aperture size:}
The size of the individual apertures of a 
ground-based interferometer is limited to the size of an atmospheric 
turbulence cell - about 30~cm at optical wavelengths. In principle, the use of
{\it Adaptive Optics} can overcome this limit, but the need for a bright
reference source (a ``guide star'') within the isoplanatic patch
severely limits the number of suitable targets.

In space, on the other hand, there is no turbulent atmosphere. Therefore,
the size of interferometer elements is limited only by our technical
and financial ability to manufacture - and bring into space - large mirrors. 

{\bf Frame time:}
Another important obstacle for interferometry
on the ground imposed by the earth's atmosphere is the need to record the
complex visibility within a coherence time of the atmosphere - about 20~ms
at a good site for optical wavelengths. Together with the above limit on
aperture size, this severely constrains the target brightness: it has 
to be bright enough to overcome the noise associated with the 
detection system, both from the detector and the read-out electronics.

Even with a perfect Adaptive Optics system, the integration time at each
\uv -point is limited because of the Earth's rotation: the
measurement has to be completed before the baseline has changed noticably
in order to avoid \uv -smearing.

In space, the available integration time for each \uv -coordinate
is in principle unlimited, so that much fainter objects can be observed.
This assumes, however, that structural vibrations or slow drifts are 
perfectly corrected which might not be the case. 

{\bf Passbands:}
Astronomy from the ground is limited to
a number of small regions of the electro-magnetic spectrum over
which the Earth's atmosphere is transparent. In space, no such
limitations exist, and important wavebands such as the ultraviolet
or the mid-infrared become accessible. 

{\bf Environment:}
Clearly, operating an interferometer in space
to the extreme precision required for successful fringe tracking, is an
extremely challenging task. In the case of single spacecraft with a number
of apertures on a connecting truss, the mechanical vibrations of the
structure must be minimized, and even for low-noise structures, residual
motions must be actively corrected. In principle, however, space is a 
favorable environment because it is intrinsically much quieter than the 
geologically active surface of the Earth. Because of this and the limitations
of a truss with regard to the achievable baseline length, future
synthesis imaging in space is likely to be realized with a suite of
free-flying apertures which are positionally controlled with respect
to each other and the beam combining optics by reference laser beams.
\subsection*{Ongoing and Planned Projects}
The technology which makes the ambitious methods of active pathlength 
control and fringe tracking feasible has only recently become available,
mostly through the development of ever faster computers which allow
real-time control of mechanical instabilities.
While the principles of synthesis imaging are well understood, and
the technique has been demonstrated on the ground,
interferometry in space has yet to be realized. 
Today, a number of space-based interferometer projects are in the 
development or planning stages as milestones of NASA's {\sc Origins Program}.
Most of these missions will be capable of performing synthesis imaging: 

{\bf Space Technology 3 (ST3)}
is intended to demonstrate the feasibility of precision control
for free-flying spacecrafts. It will consist of two spacecrafts, one
with a collector and the beam combining optics, and one that serves only
as a collector. Launch is planned for 2003. 

{\bf Space Interferometry Mission (SIM)}
consists of a single spacecraft with a 12~m truss which
contains eight telescopes, each with a diameter of about 30~cm.
SIM primary science goal is to perform high-precision astrometry
for a number of science programs. Because SIM will allow to rotate
the truss structure, and to combine any pair of two out of the eight
telescopes, it has great potential for synthesis imaging. SIM will be
launched in 2006. 

{\bf Terrestrial Planet Finder (TPF)} in its current design
comprises four free-flying
3.5~m telescopes and a fifth spacecraft with the beam-combining optics.
The design allows maximum baselines of about 1~km for a resolution of
less than 0.001\as\ at wavelengths around 3~$\mu$m.
The main scientific goal of TPF is the detection
and characterization of Earth-like planets at infrared wavelengths by
means of interferometric {\sc Nulling} of the light from the host star. 
A tentative launch date for TPF is 2010. 

{\bf DARWIN}
is a mission concept currently proposed for a cornerstone mission
of the European Space Agency (ESA) with a launch after 2009. It has similar
design concept and scientific goals as TPF.  

While the basic technological tools for these ambitious missions
are in hand, their successful implementation will occur in steps and
require time, effort, and money. On the way to the ultimate goal of
taking images of other Earths, major advances in many other aspects
of astronomy are almost certain due to the dramatic increase in spatial
resolution that only synthesis imaging can provide. 
\subsubsection*{Bibliography}
The complete treatment of wave diffraction and interference is treated
in great detail in Born, M. \& Wolf, E. 1980 {\it Principles of Optics}
(Oxford: Pergamon) 6th Ed. \\
A more detailed review about the intricacies of interferometry both
from the ground and in space can be found in Shao, M. \& Colavita, M. M.
1992, Long baseline optical and infrared stellar interferometry
{\it Ann. Rev. of Astronomy \& Astrophysics} {\bf 30} 457-498 \\
TORSTEN B\"OKER

\vfill

{\bf Fig. 1} Illustration of the principle of synthesis imaging: a) model
source to be imaged. b-d) Image reconstructions (right) for various baseline
distributions (left). The ring-like residuals around the central
source image are called grating rings. They are a consequence of the 
regularly spaced gaps in the \uv -coverage, and
can be removed to a large extent by numerical image 
restoration algorithms.

\end{document}